\documentclass[two column]{revtex4}

\newcommand{\ie}   {{\em i.e.}}
\newcommand{\half}  {\frac{1}{2}}
\renewcommand{\bar}{\overline}
\newcommand{\M}{{\cal M}}
\newcommand{\VEV}[1]{\left\langle{#1}\right\rangle}
\newcommand{\ket}[1]{\vert\,{#1}\rangle}

\begin{document}
%\preprint{SLAC-PUB-9642}

\title{Gauge Theories on the Light-Front}

\author{Stanley J. Brodsky}
\affiliation{Stanford Linear Accelerator Center \\
Stanford University, Stanford, California 94309}
\email{sjbth@slac.stanford.edu}
\date{\today}

\begin{abstract}
The light-front quantization of gauge theories in light-cone gauge
provides a frame-independent
wavefunction representation of relativistic bound states, simple
forms for current matrix elements, explicit unitary, and a trivial
vacuum.  The light-front Hamiltonian form of QCD
provides an alternative to lattice gauge theory for the computation of
nonperturbative quantities such as the hadronic spectrum and  the
corresponding eigenfunctions.  In the case of the electroweak
theory, spontaneous symmetry breaking is represented by the
appearance of zero modes of the Higgs field.  Light-front
quantization then leads to an elegant ghost-free theory of massive
gauge particles, automatically incorporating the Lorentz and 't
Hooft conditions, as well as the Goldstone boson equivalence
theorem.
\end{abstract}

\maketitle

\section{Introduction}

One of the challenges of relativistic quantum field theory is to
compute the wavefunctions of bound states such as the amplitudes
which determine the quark and gluon substructure of hadrons in
quantum chromodynamics.  In light-front
quantization~\cite{Dirac:cp}, one fixes the initial boundary
conditions of a composite system as its constituents are
intercepted by a single light-wave evaluated on the hyperplane
$x^+ = t + z/c$.  The light-front quantization of QCD  provides a
frame-independent, quantum-mechanical representation of a hadron
at the amplitude level, capable of encoding its multi-quark,
hidden-color and gluon momentum, helicity, and flavor correlations
in the form of universal process-independent and frame-independent
hadron wavefunctions~\cite{Brodsky:1997de}.  Remarkably, quantum
fluctuations of the vacuum are absent if one uses light-front time
to quantize the system, so that matrix elements such as the
electromagnetic form factors only depend on the currents of the
constituents described by the light-cone wavefunctions.  The
degrees of freedom associated with vacuum phenomena such as
spontaneous symmetry breaking in the Higgs model have their
counterpart in light-front $k^+ =0$ zero modes of the fields.

In Dirac's ``Front Form"~\cite{Dirac:cp}, the generator of
light-front time translations is $P^- = i{ \partial\over \partial
\tau}.$ Boundary conditions are set on the transverse plane
labelled by $x_\perp$ and $x^- = z-ct$.  Given the Lagrangian of a
quantum field theory, $P^-$ can be constructed as an operator on
the Fock basis, the eigenstates of the free theory.  Since each
particle in the Fock basis is on its mass shell, $k^- \equiv
k^0-k^3 = {k^2_\perp + m^2 \over k^+},$ and its energy $k^0 =\half
( k^+ + k^-) $ is positive, only particles with positive momenta
$k^+ \equiv k^0 + k^3 \ge 0$ can occur in the Fock basis.  Since
the total plus momentum $P^+ = \sum_n k^+_n$ is conserved, the
light-cone vacuum cannot have any particle content.

The Heisenberg equation on the light-front is
\begin{equation}
H_{LC} \ket{\Psi} = M^2 \ket{\Psi}\ .
\end{equation}
The operator
$H_{LC} = P^+ P^- - P^2_\perp,$ the ``light-cone Hamiltonian", is
frame-independent.
This can in principle be solved by diagonalizing the matrix
$\VEV{n|H_{LC}|m}$ on the free Fock basis:~\cite{Brodsky:1997de}
\begin{equation}
\sum_m \VEV{n|H_{LC}|m}\VEV{m|\psi} = M^2 \VEV{n|\Psi}\ .
\end{equation}
The eigenvalues $\{M^2\}$ of
$H_{LC}=H^{0}_{LC} + V_{LC}$ give the squared invariant masses of
the bound and continuum spectrum of the theory.  The light-front
Fock space is the eigenstates of the free light-front Hamiltonian;
\ie, it is a Hilbert space of non-interacting quarks and gluons,
each of which satisfy $k^2 = m^2$ and $k^- = {m^2 + k^2_\perp
\over k^+} \ge 0.$  The projections $\{\VEV{n|\Psi}\}$ of the
eigensolution on the $n$-particle Fock states provide the
light-front wavefunctions.  Thus solving a quantum field theory is
equivalent to solving a coupled many-body quantum mechanical
problem:
\begin{equation}
\left[M^2 - \sum_{i=1}^n{m^2 + k^2_\perp\over x_i}\right] \psi_n =
\sum_{n'}\int \VEV{n|V_{LC}|n'} \psi_{n'}\end{equation}
where the convolution and sum is understood over the Fock number,
transverse momenta, plus momenta, and helicity of the intermediate
states.
Light-front wavefunctions are also related to momentum-space
Bethe-Salpeter wavefunctions by integrating over the relative
momenta $k^- = k^0 - k^z$ since this projects out the dynamics at
$x^+ =0.$

The light-front quantization of gauge theory can be most
conveniently carried out in the light-cone gauge $A^+ = A^0 + A^z
= 0$.  In this gauge the $A^-$ field becomes a dependent degree of
freedom, and it can be eliminated from the Hamiltonian in favor of
a set of specific instantaneous light-front time interactions.  In
fact in $QCD(1+1)$ theory, this instantaneous interaction provides
the confining linear $x^-$ interaction between quarks.  In $3+1$
dimensions, the transverse field $A^\perp$ propagates massless
spin-one gluon quanta with polarization vectors
\cite{Lepage:1980fj} which satisfy both the gauge condition
$\epsilon^+_\lambda = 0$ and the Lorentz condition $k\cdot
\epsilon= 0$.

LF quantization is especially useful for quantum chromodynamics,
since it provides a rigorous extension of many-body quantum
mechanics to relativistic bound states:  the quark, and gluon
momenta and spin correlations of a hadron become encoded in the
form of universal process-independent, Lorentz-invariant
wavefunctions \cite{bro}.
For example, the eigensolution of a
meson in QCD,  projected on the eigenstates $\{\ket{n} \}$ of the free
Hamiltonian $ H^{QCD}_{LC}(g = 0)$ at fixed $\tau = t-z/c$ with the
same global quantum numbers, has the expansion:
\begin{eqnarray}
&&\left\vert \Psi_M; P^+, {\vec P_\perp}, \lambda \right> =
\nonumber
\\ && \sum_{n \ge 2,\lambda_i} \int \Pi^{n}_{i=1} {d^2k_{\perp i}
dx_i \over \sqrt{x_i} 16 \pi^3}\nonumber\\ &&
 \times 16 \pi^3 \delta\left(1- \sum^n_j x_j\right) \delta^{(2)}
\left(\sum^n_\ell \vec k_{\perp \ell}\right) \\[1ex]
&&\times \left\vert n; x_i P^+, x_i {\vec P_\perp} + {\vec
k_{\perp i}}, \lambda_i\right
> \psi_{n/M}(x_i,{\vec k_{\perp i}},\lambda_i)  .\nonumber
\end{eqnarray}
The set of light-front Fock state wavefunctions $\{\psi_{n/M}\}$
represent the ensemble of quark and gluon states possible when the
meson is intercepted at the light-front.  The light-front momentum
fractions $x_i = k^+_i/P^+_\pi = (k^0 + k^z_i)/(P^0+P^z)$ with
$\sum^n_{i=1} x_i = 1$ and ${\vec k_{\perp i}}$ with $\sum^n_{i=1}
{\vec k_{\perp i}} = {\vec 0_\perp}$ represent the relative
momentum coordinates of the QCD constituents and are independent
of the total momentum of the state.

Remarkably, the scalar light-front wavefunctions
$\psi_{n/p}(x_i,{\vec k_{\perp i}},\lambda_i)$ are independent of
the proton's momentum $P^+ = P^0 + P^z$, and $P_\perp$.  (The
light-cone spinors and polarization vectors multiplying
$\psi_{n/p}$ are functions of the absolute coordinates.) Thus once
one has solved for the light-front wavefunctions, one can compute
hadron matrix elements of currents between hadronic states of
arbitrary momentum.  The actual physical transverse momenta are
${\vec p_{\perp i}} = x_i {\vec P_\perp} + {\vec k_{\perp i}}.$
The $\lambda_i$ label the light-front spin $S^z$ projections of
the quarks and gluons along the quantization $z$ direction.  The
spinors of the light-front formalism automatically incorporate the
Melosh-Wigner rotation.  The physical gluon polarization vectors
$\epsilon^\mu(k,\ \lambda = \pm 1)$ are specified in light-cone
gauge by the conditions $k \cdot \epsilon = 0,\ \eta \cdot
\epsilon = \epsilon^+ = 0.$ The parton degrees of freedom are thus
all physical; there are no ghost or negative metric states.

\section{Properties of Light-Front Wavefunctions}

An important feature of the light-front formalism is that the
projection $J_z$ of the total is kinematical and conserved.  Each
light-front Fock state component thus satisfies the angular
momentum sum rule: $ J^z = \sum^n_{i=1} S^z_i + \sum^{n-1}_{j=1}
l^z_j \ . $ The summation over $S^z_i$ represents the contribution
of the intrinsic spins of the $n$ Fock state constituents.  The
summation over orbital angular momenta
\begin{equation}
l^z_j = -{\mathrm i} \left(k^1_j\frac{\partial}{\partial k^2_j}
-k^2_j\frac{\partial}{\partial k^1_j}\right)
\end{equation}
derives from the $n-1$ relative momenta.  This excludes the
contribution to the orbital angular momentum due to the motion of
the center of mass, which is not an intrinsic property of the
hadron.  The numerator structure of the light-front wavefunctions
is in large part determined by the angular momentum constraints.
Thus wavefunctions generated by perturbation theory provide
a template for the numerator structure of nonperturbative light-front
wavefunctions.

Dae Sung Hwang, Bo-Qiang Ma, Ivan Schmidt, and I
\cite{Brodsky:2001ii} have  shown that the light-front
wavefunctions generated by the radiative corrections to the
electron in QED provide a simple system for understanding the spin
and angular momentum decomposition of relativistic systems.  This
perturbative model also illustrates the interconnections between
Fock states of different particle number.  The model is patterned
after the quantum structure which occurs in the one-loop Schwinger
${\alpha / 2 \pi} $ correction to the electron magnetic moment
\cite{Brodsky:1980zm}.  In effect, we can represent a spin-$\half$
~ system as a composite of a spin-$\half$ ~ fermion and spin-one
vector boson.  A similar model has been used to illustrate the
matrix elements and evolution of light-front helicity and orbital
angular momentum operators \cite{Harindranath:1999ve}.  This
representation of a composite system is particularly useful
because it is based on two constituents but yet is totally
relativistic.  We can then explicitly compute the form factors
$F_1(q^2)$ and $F_2(q^2)$ of the electromagnetic current and the
various contributions to the form factors $A(q^2)$ and $B(q^2)$ of
the energy-momentum tensor.

Recently Ji, Ma, and Yuan~\cite{Ji:2003fw} have derived perturbative QCD
counting rules for light-front wavefunctions with general values of
orbital angular momentum which constrain their form at large transverse
momentum.

\section{Applications of Light-Front Wavefunctions}

Matrix elements of spacelike currents such as spacelike
electromagnetic form factors have an exact representation in terms
of simple overlaps of the light-front wavefunctions in momentum
space with the same $x_i$ and unchanged parton
number $n$~\cite{Drell:1970km,West:1970av,Brodsky:1980zm}.  The Pauli
form factor and anomalous moment are spin-flip matrix elements of
$j^+$  and thus connect states with $\Delta L_z =1.$  Thus, these
quantities are nonzero only if there is nonzero orbital angular
momentum of the quarks in the proton.   The Dirac form factor is
diagonal in $L_z$ and is typically dominated at high $Q^2$ by highest
states with the highest orbital angular momentum.

The formulas for electroweak current matrix elements of $j^+$ can be
easily extended to the $T^{++}$ coupling of gravitons.  In, fact, one can
show that the anomalous gravito-magnetic moment $B(0)$, analogous to
$F_2(0)$ in electromagnetic current interactions, vanishes identically
for any system, composite or elementary~\cite{Brodsky:2001ii}.  This
important feature, which follows in general from the equivalence
principle~\cite{Okun,Ji:1996kb,Ji:1997ek,Ji:1997nm,Teryaev:1999su},
is obeyed explicitly in the light-front
formalism~\cite{Brodsky:2001ii}.

The light-front Fock
representation is specially advantageous in the study of exclusive
$B$ decays.  For example, we can write down an exact
frame-independent representation of decay matrix elements such as
$B \to D \ell \bar \nu$ from the overlap of $n' = n$ parton
conserving wavefunctions and the overlap of $n' = n-2$ from the
annihilation of a quark-antiquark pair in the initial wavefunction
\cite{Brodsky:1999hn}.  The off-diagonal $n+1 \rightarrow n-1$
contributions give a new perspective for the physics of
$B$-decays.  A semileptonic decay involves not only matrix
elements where a quark changes flavor, but also a contribution
where the leptonic pair is created from the annihilation of a $q
{\bar{q'}}$ pair within the Fock states of the initial $B$
wavefunction.  The semileptonic decay thus can occur from the
annihilation of a nonvalence quark-antiquark pair in the initial
hadron.

The ``handbag" contribution to the leading-twist
off-forward parton distributions measured in deeply virtual
Compton scattering has a similar light-front wavefunction
representation as overlap integrals of light-front wavefunctions
\cite{Brodsky:2000xy,Diehl:2000xz}.

The distribution amplitudes $\phi(x_i,Q)$ which appear in
factorization formulae for hard exclusive processes are the
valence LF Fock wavefunctions integrated over the relative
transverse momenta up to the resolution scale
$Q$~\cite{Lepage:1980fj}.  These quantities specify how a hadron
shares its longitudinal momentum among its valence quarks; they
control virtually all exclusive processes involving a hard scale
$Q$, including form factors, Compton scattering and
photoproduction at large momentum transfer, as well as the decay
of a heavy hadron into specific final states
\cite{Beneke:1999br,Keum:2000ph}.

The quark and gluon probability distributions $q_i(x,Q)$ and
$g(x,Q)$ of a hadron can be computed from the absolute squares of
the light-front wavefunctions, integrated over the transverse
momentum.  All helicity distributions are thus encoded in terms of
the light-front wavefunctions.  The DGLAP evolution of the
structure functions can be derived from the high $k_\perp$
properties of the light-front wavefunctions.  Thus given the
light-front wavefunctions, one can compute \cite{Lepage:1980fj}
all of the leading twist helicity and transversity distributions
measured in polarized deep inelastic lepton scattering.
Similarly, the transversity distributions and off-diagonal
helicity convolutions are defined as a density matrix of the
light-front wavefunctions.

However, it is not true that the leading-twist structure functions
$F_i(x,Q^2)$  measured in deep inelastic lepton scattering are
identical to the quark and gluon distributions.  For example, it
is usually assumed, following the parton model,  that the $F_2$
structure function measured in neutral current deep inelastic
lepton scattering is at leading order in $1/Q^2$ simply
$F_2(x,Q^2) =\sum_q  e^2_q  x q(x,Q^2)$, where $x = x_{bj} = Q^2/2
p\cdot q$ and $q(x,Q)$ can be computed from the absolute square of
the proton's light-front wavefunction.  Recent work by Hoyer,
Marchal,  Peigne, Sannino, and myself shows that this standard
identification is wrong~\cite{Brodsky:2002ue}. Gluon exchange
between the fast, outgoing partons and the target spectators,
which is usually assumed to be an irrelevant gauge artifact,
actually affects the leading-twist structure functions in a
profound way.   The diffractive scattering of the fast outgoing
quarks on spectators in the target in turn causes shadowing in the
DIS cross section. Thus the depletion of the nuclear structure
functions is not intrinsic to the wave function of the nucleus,
but is a coherent effect arising from the destructive interference
of diffractive channels induced by final-state interactions.  Thus
the shadowing corrections related to the Gribov-Glauber mechanism,
the interference effects of leading twist diffractive processes in
nuclei are separate effects in deep inelastic scattering, are  not
computable from the bound state wavefunctions of the target
nucleon or nucleus.  Similarly, the effective pomeron distribution
of a hadron is not derived from its light-front wavefunction and
thus is not a universal property.

Measurements from the HERMES and SMC collaborations show a
remarkably large single-spin asymmetry in semi-inclusive pion
leptoproduction $\gamma^*(q) p \to \pi X$ when the proton is
polarized normal to the photon-to-pion production plane.  Recently,
Hwang, Schmidt, and I~\cite{Brodsky:2002cx} have shown that
final-state interactions from gluon exchange between the outgoing
quark and the target spectator system lead to single-spin
asymmetries in deep inelastic lepton-proton scattering at leading
twist in perturbative QCD; {\it i.e.}, the rescattering
corrections are not power-law suppressed at large photon
virtuality $Q^2$ at fixed $x_{bj}$.  The existence of such
single-spin asymmetries requires a phase difference between two
amplitudes coupling the proton target with $J^z_p = \pm {1\over
2}$ to the same final-state, the same amplitudes which are
necessary to produce a nonzero proton anomalous magnetic moment.
The single-spin asymmetry which arises from such final-state
interactions does not factorize into a product of distribution
function and fragmentation function, and it is not related to the
transversity distribution $\delta q(x,Q)$ which correlates
transversely polarized quarks with the spin of the transversely
polarized target nucleon.  These effects highlight the unexpected
importance of final- and initial-state interactions in QCD
observables---they lead to leading-twist single-spin asymmetries,
diffraction, and nuclear shadowing, phenomena not included in the
wavefunction of the target.

\section{Measurements of Light-Front Wavefunctions}

It is possible to measure the light-front wavefunctions of a
relativistic hadron  by diffractively dissociating it into jets in
high-energy hadron-nucleus collisions such as $\pi A \to {\rm jet
jet A^\prime}.$ Only the configurations of the incident hadron
which have small transverse size and minimal color dipole moments
can traverse the nucleus with minimal interactions and leave it
intact.   The forward diffractive amplitude is thus coherent over
the entire nuclear volume and  proportional to nuclear number $A$.
The fractional momentum distribution of the jets is correlated
with the valence quarks' light-cone momentum fractions $x_i.$
\cite{Ashery:1999nq,Bertsch:1981py,Frankfurt:1993it,Frankfurt:2000tq}.
The QCD mechanisms for hard diffractive dissociation can be more
complicated in the case of proton targets. A review and references
is given in Ref.~\cite{Brodsky:2002st}.

The fact that  Fock states of a hadron with small particle number
and small impact separation have small color dipole moments and
weak hadronic interactions is a remarkable manifestation of the
gauge structure of QCD.  It is the basis for the predictions for
``color transparency" in hard quasi-exclusive
\cite{Brodsky:1988xz,Frankfurt:1988nt} and diffractive reactions
\cite{Bertsch:1981py,Frankfurt:1993it,Frankfurt:2000tq}. The E791
experiment at FermiLab has verified the nuclear number scaling
predictions and  have thus provided a remarkable confirmation of
this consequence of QCD color transparency~\cite{Ashery:1999nq}.
The new EVA spectrometer experiment E850 at Brookhaven has also
reported striking effects of color transparency in quasi-elastic
proton-proton scattering in nuclei~\cite{Leksanov:2001ui}.

The CLEO collaboration~\cite{Gronberg:1997fj}  has verified the
scaling and angular predictions for the photon-meson to meson form
factor $F_{\gamma \pi^0}(q^2)$ which is measured in $\gamma^*
\gamma \to \pi^0$ reactions. The  results are in close agreement
with the scaling and normalization of the PQCD
predictions~\cite{Brodsky:1981rp}, provided that the pion
distribution amplitude $\phi_\pi(x,Q)$ is close to the $x(1-x)$
form, the asymptotic solution to the evolution equation.  The pion
light-front momentum distribution measured in diffractive dijet
production in pion-nucleus collisions by the E791
experiment~\cite{Ashery:1999nq} has a similar
form~\cite{Ashery:2003cd}. Data~\cite{Aihara:qk} for $\gamma
\gamma \to \pi^+ \pi^+ + K^+ K^-$ at $W = \sqrt s > 2. 5$ GeV
are also in  agreement with the perturbative QCD predictions.
Moreover, the angular distribution shows the expected transition
to the predicted QCD form as $W$ is raised.   A compilation of the
two-photon data has been given by Whalley~\cite{Whalley:2001mk}.
Measurements of the reaction $\gamma \gamma \to \pi^0 \pi^0$ are
highly sensitive to the shape of the pion distribution amplitude.
The perturbative QCD predictions~\cite{Brodsky:1981rp} for this
channel contrast strongly with model predictions based on the QCD
Compton handbag diagram~\cite{Diehl:2001fv}.

\section{Higher Particle-Number Fock States}

The light-front Fock state expansion of a hadron contains
fluctuations with an arbitrary number of quark and gluon partons.
The higher Fock states of the light hadrons describe the sea quark
structure of the deep inelastic structure functions, including
``intrinsic" strangeness and charm fluctuations specific to the
hadron's structure rather than gluon substructure
\cite{Brodsky:1980pb,Brodsky:1981se}.  The maximal contribution of
an intrinsic heavy quark occurs at $x_Q \simeq {m_{\perp Q}/
\sum_i m_\perp}$ where $m_\perp = \sqrt{m^2+k^2_\perp}$; \ie\ at
large $x_Q$, since this minimizes the invariant mass $\M^2_n$.
The measurements of the charm structure function by the EMC
experiment are consistent with intrinsic charm at large $x$ in the
nucleon with a probability of order $0.6 \pm 0.3 \% $
\cite{Harris:1996jx} which is consistent with the recent estimates
based on instanton fluctuations \cite{Franz:2000ee}.  Franz,
Polyakov, and Goeke have analyzed the properties of the intrinsic
heavy-quark fluctuations in hadrons using the operator-product
expansion~\cite{Franz:2000ee}.  For example, the light-cone
momentum fraction carried by intrinsic heavy quarks in the proton
$x_{Q \bar Q}$ as measured by the $T^{+ + }$ component of the
energy-momentum tensor is related in the heavy-quark limit to the
forward matrix element $\langle p \vert {\hbox{tr}_c}
{(G^{+\alpha} G^{+ \beta} G_{\alpha \beta})/ m_Q^2 }\vert p
\rangle ,$ where $G^{\mu \nu}$ is the gauge field strength tensor.
Diagrammatically, this can be described as a heavy quark loop in
the proton self-energy with four gluons attached to the light,
valence quarks.  Since the non-Abelian commutator $[A_\alpha,
A_\beta]$ is involved, the heavy quark pairs in the proton
wavefunction are necessarily in a color-octet state.  It follows
from dimensional analysis that the momentum fraction carried by
the $Q\bar Q$ pair scales as $k^2_\perp / m^2_Q$ where $k_\perp$
is the typical momentum in the hadron wave function.  In contrast,
in the case of Abelian theories, the contribution of an intrinsic,
heavy lepton pair to the bound state's structure first appears in
${ O}(1/m_L^4)$.

The presence of intrinsic charm quarks in the $B$ wave function
provides new mechanisms for $B$ decays.  For example, Chang and Hou
have considered the production of final states with three charmed
quarks such as $B \to J/\psi D \pi$ and $B \to J/\psi
D^*$~\cite{Chang:2001iy}; these final states are difficult to
realize in the valence model, yet they occur naturally when the
$b$ quark of the intrinsic charm Fock state $\ket{ b \bar u c \bar
c}$ decays via $b \to c \bar u d$.  Susan Gardner and I have shown
that the presence of intrinsic charm in the hadrons' light-front
wave functions, even at a few percent level, provides new,
competitive decay mechanisms for $B$ decays which are nominally
CKM-suppressed~\cite{Brodsky:2001yt}.  Similarly, Karliner and I
\cite{Brodsky:1997fj} have shown that the transition $J/\psi \to
\rho \pi$ can occur by the rearrangement of the $c \bar c$ from
the $J/\psi$ into the $\ket{ q \bar q c \bar c}$ intrinsic charm
Fock state of the $\rho$ or $\pi$.  On the other hand, the overlap
rearrangement integral in the decay $\psi^\prime \to \rho \pi$
will be suppressed since the intrinsic charm Fock state radial
wavefunction of the light hadrons will evidently not have nodes in
its radial wavefunction.  This observation provides a natural
explanation of the long-standing puzzle~\cite{Brodsky:1987bb} why
the $J/\psi$ decays prominently to two-body pseudoscalar-vector
final states, breaking hadron helicity
conservation~\cite{Brodsky:1981kj}, whereas the $\psi^\prime$ does
not.

\section{Light-Front Quantization of QCD}

Quantum field theories are usually quantized at fixed ``instant"
time $t$.  The resulting Hamiltonian theory is complicated by the
dynamical nature of the vacuum state and the fact that
relativistic boosts are not kinematical but involve interactions.
The calculation of even the simplest current matrix elements
requires the computation of amplitudes where the current interacts
with particles resulting from the fluctuations of the vacuum.  All
of these problems are dramatically alleviated when one quantizes
quantum field theories at fixed light-cone time $\tau.$  A review
of the development of light-front quantization of QCD and other
quantum field theories is given in Ref.~\cite{Brodsky:1997de}.

Prem Srivastava and I \cite{Srivastava:2000cf} have  presented a
new systematic study of light-front-quantized gauge theory in
light-cone gauge using a Dyson-Wick S-matrix expansion based on
light-front-time-ordered products.  The Dirac bracket method is used to
identify the independent field degrees of freedom~\cite{Ditman}.  In our
analysis one imposes the light-cone gauge condition as a linear
constraint using a Lagrange multiplier, rather than a quadratic form.  We
then find that the LF-quantized free gauge theory simultaneously
satisfies the covariant gauge condition
$\partial\cdot A=0$ as an operator condition as well as the LC gauge
condition.  The gluon propagator has the form
\begin{equation}
\VEV{0|\,T({A^{a}}_{\mu}(x){A^{b}}_{\nu}(0))\,|0} ={{i\delta^{ab}}
\over {(2\pi)^{4}}} \int d^{4}k \;e^{-ik\cdot x} \; \;
{D_{\mu\nu}(k)\over {k^{2}+i\epsilon}}
\end{equation}
where we have defined
\begin{equation}
D_{\mu\nu}(k)= D_{\nu\mu}(k)= -g_{\mu\nu} + \frac
{n_{\mu}k_{\nu}+n_{\nu}k_{\mu}}{(n\cdot k)} - \frac {k^{2}}
{(n\cdot k)^{2}} \, n_{\mu}n_{\nu}.
\end{equation}
Here $n_{\mu}$ is a null four-vector, gauge direction, whose
components are chosen to be $\, n_{\mu}={\delta_{\mu}}^{+}$, $\,
n^{\mu}={\delta^{\mu}}_{-}$.  Note also
\begin{eqnarray}
D_{\mu\lambda}(k) {D^{\lambda}}_{\nu}(k)=
D_{\mu\perp}(k) {D^{\perp}}_{\nu}(k)&=& - D_{\mu\nu}(k),  \\
 k^{\mu}D_{\mu\nu}(k)=0,  n^{\mu}D_{\mu\nu}(k)&\equiv&
D_{-\nu}(k)=0, \nonumber \\ D_{\lambda\mu}(q) \,D^{\mu\nu}(k)\,
D_{\nu\rho}(q') &=& -D_{\lambda\mu}(q)D^{\mu\rho}(q').\nonumber
\end{eqnarray}
The gauge field propagator $\,\,i\,D_{\mu\nu}(k)/
(k^{2}+i\epsilon)\,$ is transverse not only to the gauge direction
$n_{\mu}$ but also to $k_{\mu}$, {\em i.e.}, it is {\it
doubly-transverse}. Thus $D$ represents the polarization sum over
physical propagating modes.  The last term proportional to $n_\mu
n_\nu$ in the gauge propagator does not  appear in the usual
formulations of light-cone gauge. However, in tree graph
calculations it cancels against instantaneous gluon exchange
contributions.

The remarkable properties of (the projector) $D_{\nu\mu}$ greatly
simplifies the computations of loop amplitudes.  For example, the
coupling of gluons to propagators carrying high momenta is
automatic.  In the case of tree graphs, the term proportional to
$n_{\mu}n_{\nu}$ cancels against the instantaneous gluon exchange
term.  However, in the case of loop diagrams, the separation needs
to be maintained so that one can identify the correct
one-particle-irreducible contributions.   The absence of collinear
divergences in irreducible diagrams in the light-cone gauge
greatly simplifies the leading-twist factorization of soft and
hard gluonic corrections in high momentum transfer inclusive and
exclusive reactions \cite{Lepage:1980fj} since the numerators
associated with the gluon coupling only have transverse
components.

The interaction Hamiltonian of QCD in light-cone gauge can be
derived by systematically applying the Dirac bracket method to
identify the independent fields \cite{Srivastava:2000cf}.  It
contains the usual Dirac interactions between the quarks and
gluons, the three-point and four-point gluon non-Abelian
interactions plus instantaneous light-front-time gluon exchange
and quark exchange contributions
\begin{eqnarray}
{\cal H}_{int}&=&
  -g \,{{\bar\psi}}^{i}
\gamma^{\mu}{A_{\mu}}^{ij}{{\psi}}^{j}   \nonumber \\
&& +\frac{g}{2}\, f^{abc} \,(\partial_{\mu}{A^{a}}_{\nu}-
\partial_{\nu}{A^{a}}_{\mu}) A^{b\mu} A^{c\nu} \nonumber \\
&& +\frac {g^2}{4}\,
f^{abc}f^{ade} {A_{b\mu}} {A^{d\mu}} A_{c\nu} A^{e\nu} \nonumber \\
&& - \frac{g^{2}}{ 2}\,\, {{\bar\psi}}^{i} \gamma^{+}
\,(\gamma^{\perp'}{A_{\perp'}})^{ij}\,\frac{1}{i\partial_{-}} \,
(\gamma^{\perp} {A_{\perp}})^{jk}\,{\psi}^{k} \nonumber \\
&& -\frac{g^{2}}{ 2}\,{j^{+}}_{a}\, \frac
{1}{(\partial_{-})^{2}}\, {j^{+}}_{a}
\end{eqnarray}
where
\begin{equation}
{j^{+}}_{a}={{\bar\psi}}^{i} \gamma^{+} (
{t_{a}})^{ij}{{\psi}}^{j} + f_{abc} (\partial_{-} A_{b\mu})
A^{c\mu} \ .
\end{equation}

The renormalization constants in the non-Abelian theory have been
shown~\cite{Srivastava:2000cf} to satisfy the identity $Z_1=Z_3$
at one-loop order, as expected in a theory with only physical
gauge degrees of freedom.  The renormalization factors in the
light-cone gauge are independent of the reference direction
$n^\mu$.   The QCD $\beta$ function computed in the noncovariant
LC gauge agrees with the conventional theory
result~\cite{gross,polit}.  Dimensional regularization and the
Mandelstam-Leibbrandt
prescription~\cite{Mandelstam:1982cb,Leibbrandt:1987qv,Bassetto:1984dq}
for LC gauge were used to define the Feynman loop
integrations~\cite{Bassetto:1996ph}.  There are no Faddeev-Popov
or Gupta-Bleuler ghost terms.

The running coupling constant and QCD $\beta$ function have also
been computed at one loop in the doubly-transverse light-cone
gauge \cite{Srivastava:2000cf}.  It is also possible to
effectively quantize QCD using light-front methods in covariant
Feynman gauge \cite{Srivastava:2000gi}.

It is well-known that the light-cone gauge itself is not
completely defined until one specifies a prescription for the
poles of the gauge propagator at $n \cdot k= 0.$ The
Mandelstam-Liebbrandt prescription has the advantage of preserving
causality and analyticity, as well as leading to proofs of the
renormalizability and unitarity of Yang-Mills
theories~\cite{Bassetto:1991ue}.  The ghosts which appear in
association with the Mandelstam-Liebbrandt prescription from the
single poles have vanishing residue in absorptive parts, and thus
do not disturb the unitarity of the theory.

A remarkable advantage of light-front quantization is that the
vacuum state $\ket{0}$ of the full QCD Hamiltonian evidently
coincides with the free vacuum.  The light-front vacuum is
effectively trivial if the interaction Hamiltonian applied to the
perturbative vacuum is zero.  Note that all particles in the
Hilbert space have positive energy $k^0 = {1\over 2}(k^+ + k^-)$,
and thus positive light-front $k^\pm$.  Since the plus momenta
$\sum k^+_i$ is conserved by the interactions, the perturbative
vacuum can only couple to states with particles in which all
$k^+_i$ = 0; \ie, so called zero-mode states.   Bassetto and
collaborators \cite{Bassetto:1999tm} have shown that the computation of
the spectrum of $QCD(1+1)$ in equal time quantization requires
constructing the full spectrum of non perturbative contributions
(instantons).  In contrast, in the light-front quantization of gauge
theory, where the $k^+ = 0 $ singularity of the instantaneous interaction
is defined by a simple infrared regularization, one obtains the correct
spectrum of $QCD(1+1)$ without any need for vacuum-related contributions.

Zero modes of auxiliary fields are necessary to distinguish the
theta-vacua of massless QED(1+1)
\cite{Yamawaki:1998cy,McCartor:2000yy,Srivastava:1999et}, or to
represent a theory in the presence of static external boundary
conditions or other constraints.  Zero-modes provide the
light-front representation of spontaneous symmetry breaking in
scalar theories \cite{Pinsky:1994yi}.

\section{Light-Front Quantization of the Standard Model}

Prem Srivastava and I have also shown how light-front quantization can
be applied to the Glashow, Weinberg and Salam (GWS) model of
electroweak interactions based on the nonabelian gauge group
$SU(2)_{W}\times U(1)_{Y}$~\cite{gws}.   This theory contains a
nonabelian Higgs sector which triggers spontaneous symmetry
breaking (SSB).  A convenient way of implementing SSB and the
(tree level) Higgs mechanism in the {\it front form} theory  was
developed earlier by Srivastava~\cite{pre4,pre5,pre6}.  One
separates the quantum fluctuation fields from the corresponding
{\it dynamical bosonic condensate } (or
zero-longitudinal-momentum-mode) variables, {\it before} applying
the Dirac procedure in order to construct the Hamiltonian
formulation.  The canonical quantization of LC gauge GWS
electroweak theory in the {\it front form} can  be derived by
using the Dirac procedure to construct a self-consistent LF
Hamiltonian theory.  This leads to an attractive new formulation
of the Standard Model of the strong and electroweak interactions
which does not break the physical vacuum and has well-controlled
ultraviolet behavior.  The only ghosts which appear in the
formalism are the $n \cdot k = 0$ modes of the gauge field
associated with regulating the light-cone gauge prescription.  The
massive gauge field propagator has good asymptotic behavior in
accordance with a renormalizable theory, and the massive would-be
Goldstone fields can be taken as physical degrees of freedom.

For example, consider the Abelian Higgs model.  The interaction
Lagrangian is
\begin{equation} {\cal L}= -{1\over 4} F_{\mu
\nu}F^{\mu \nu}+ \vert D_\mu \phi\vert^2 -V(\phi^\dagger \phi)
\end{equation}
 where
\begin{equation}
D_\mu = \partial_\mu + i e A_\mu,\end{equation}  and
\begin{equation}V(\phi)= \mu^2 \phi^\dagger \phi + \lambda(\phi^\dagger
\phi)^2,\end{equation}
 with $\mu^2 < 0, \lambda >0.$ The complex
scalar field $\phi$ is decomposed as \begin{equation}\phi(x)=
{1\over \sqrt 2} v + \varphi(x) = {1\over \sqrt 2}[ v + h(x) + i
\eta(x)]\end{equation}
 where $v$ is the $k^+=0$ zero mode
determined by the minimum of the potential: $v^2 = -{\mu^2\over
\lambda}$, $h(x)$ is the dynamical Higgs field, and $\eta(x)$ is
the Nambu-Goldstone field.  The quantization procedure determines
$\partial \cdot A = MR$, the 't Hooft condition. One can now
eliminate the zero mode component of the Higgs field $v$ which
gives masses for the fundamental quantized fields.   The $A_\perp$
field then has mass $M=e v$ and the Higgs field acquires mass
$m^2_h = 2 \lambda v^2 = - 2 \mu^2.$

A new aspect of LF quantization, is that the third polarization of
the quantized massive vector field $A^\mu$ with four momentum
$k^\mu$ has the form $E^{(3)}_\mu = {n_\mu M / n \cdot k}$.  Since
$n^2 = 0$, this non-transverse polarization vector has zero norm.
However, when one includes the constrained interactions of the
Goldstone particle, the effective longitudinal polarization vector
of a produced vector particle is $E^{(3)}_{\rm eff \, \mu}=
E^{(3)}_\mu - { k_\mu \,k \cdot E^{(3)} / k^2}$ which is identical
to the usual polarization vector of a massive vector with norm
$E^{(3)}_{\rm eff }\cdot E^{(3)}_{\rm eff }= -1$.  Thus, unlike
the conventional quantization of the Standard Model, the Goldstone
particle only provides part of the physical longitudinal mode of
the electroweak particles.

In the LC gauge LF framework, the free massive gauge fields in the
electroweak theory satisfy simultaneously the 't Hooft conditions
as an operator equation. The sum over the three physical
polarizations is given by $K_{\mu\nu}$
\begin{eqnarray} K_{\mu\nu}(k)&=&\,\sum_{(\alpha)}
E^{(\alpha)}_{\mu}E^{(\alpha)}_{\nu} =\,D_{\mu\nu}(k)+
\frac{M^{2}}{(k^{+})^{2}}\, n_{\mu} n_{\nu}  \\
&=&-g_{\mu\nu} + \frac {n_{\mu}k_{\nu}+n_{\nu}k_{\mu}}{(n\cdot k)}
- \frac {(k^{2}-M^{2})} {(n\cdot k)^{2}} \,
n_{\mu}n_{\nu}\nonumber
\end{eqnarray} which satisfies:  $ k^{\mu}\,K_{\mu\nu}(k)=
(M^{2}/k^{+})\, n_{\nu}$ and $\,\, k^{\mu}\,
k^{\nu}\,K_{\mu\nu}(k)= M^{2}$. The free propagator of the massive
gauge field $A_{\mu}$ is
\begin{eqnarray}
&&\VEV{0\vert T \left(A_{\mu}(x)A_{\nu}(y)\right)\vert
0}= \\
&&\qquad\qquad \frac {i}{(2\pi)^{4}}\int d^{4}k \frac
{K_{\mu\nu}(k)}{ (k^{2}-M^{2}+i\epsilon)} \, e^{-i \, k\cdot
(x-y)}.\nonumber
\end{eqnarray}
It does not have the bad high energy behavior found
in the (Proca) propagator in the unitary gauge formulation, where
the would-be Nambu-Goldstone boson is gauged away.

In the limit of vanishing mass of the vector boson, the gauge
field propagator goes over to the doubly transverse gauge,
($n^{\mu}\, D_{\mu\nu}(k)=k^{\mu}\, D_{\mu\nu}(k)=0$), the
propagator found \cite{Srivastava:2000cf} in QCD.  The numerator
of the gauge propagator $K_{\mu\nu}(k)$ also has important
simplifying properties, similar to the ones associated with the
projector $D_{\mu\nu}(k)$.  The transverse polarization vectors
for massive or massless vector boson may be taken to be
$E^{\mu}_{(\perp)}(k)\equiv -D^{\mu}_{\perp}(k),$ whereas the
non-transverse third one in the massive case is found to be
parallel to the LC gauge direction $\, E^{(3)}_{\mu}(k)=
-(M/k^{+})\, n_{\mu}$.  Its projection along the direction
transverse to $k_{\mu}$ shares the spacelike vector property
carried by $E^{\mu}_{(\perp)}(k)$.  The Goldstone boson or
electroweak equivalence theorem becomes transparent in  the LF
formulation.

The interaction Hamiltonian for the Abelian Higgs model in LC gauge
$A^{+}=0$, is found to be
\begin{eqnarray}
&-&{\cal H}_{int}={\cal L}_{int} \nonumber \\ &=& e\,M\,
\,A_{\mu}A^{\mu}\,h - \frac{e\, m_{h}^{2}}{2\,M}\,(\eta^{2}+
h^{2})\, h \nonumber \\
&+& e  (h \,\partial_{\mu}\eta-\eta \,\partial_{\mu} h)\, A^{\mu}
+\frac {e^{2}}{2} ( h^{2}+\eta^{2})\,A_{\mu}A^{\mu}\,\nonumber
\\ &-& \frac{\lambda}{4}\, (\eta^{2}+ h^{2})^{2}- \frac {e^{2}}{2}\,
\,\, \left(\frac{1}{\partial_{-} \,} j^{+}\right)\,
\left(\frac{1}{\partial_{-} \,}j^{+}\right)
\end{eqnarray}
where $j_{\mu}= (h \,\partial_{\mu}\eta-\eta
\partial_{\mu}\, h)$.  The last term here is the additional quartic
instantaneous interaction in the LF theory quantized in the LC
gauge No new instantaneous cubic interaction terms are introduced.
The massive gauge field, when the mass is generated by the Higgs
mechanism, is described in our LC gauge framework by the
independent fields $A_{\perp}$ and $\eta$; the component $A^{-}$
is a dependent field.

The interaction Hamiltonian of the Standard Model can be written
in a compact form by retaining the dependent components $A^{-}$
and $\psi_{-}$ in the formulation.  Its form closely resembles the
interaction Hamiltonian of covariant theory, except for the
presence of additional instantaneous four-point interactions.  The
resulting Dyson-Wick perturbation theory expansion based on
equal-LF-time ordering has also been constructed, allowing one to
perform higher-order computations in a straightforward fashion.

The singularities in the noncovariant pieces of the field
propagators may be defined using the causal ML prescription for
$1/k^{+}$ when we employ dimensional regularization, as was shown
in our earlier work on QCD.  The power-counting rules in LC gauge
then become similar to those found in covariant gauge theory.

Spontaneous symmetry breaking is thus implemented in a novel way
when one quantizes the Standard Model at fixed light-front time
$\tau = x^+.$ In the general case, the Higgs field $\phi_i(x)$ can
be separated into two components:
\begin{equation} \phi_i(\tau,x^-,\vec x_\perp) = \omega_i(\tau,\vec x_\perp)
+ \varphi(\tau,x^-,\vec x_\perp),
\end{equation}
where $\omega_i$ is a classical $k^+=0$   zero-mode field and
$\varphi$ is the dynamical quantized field.   Here $i$ is the
weak-isospin index. The zero-mode component is determined  by
solving the Euler-Lagrange tree-level condition:
\begin{equation}
V_i^\prime(\omega) - \partial_\perp
\partial_\perp \omega_i= 0.
\end{equation}
A nonzero value for $\omega_i$ corresponds to spontaneous symmetry
breaking.  The nonzero $\omega_i$ couples to the  gauge boson and
Fermi fields through the Yukawa interactions of the Standard
Model.  It can then  be eliminated from the theory in favor of
mass terms for the fundamental matter fields in the effective
theory. The resulting masses are identical to those of the usual
Higgs implementation of spontaneous symmetry breaking in the
Standard Model.

The generators of isospin rotations are defined from the dynamical
Higgs fields:
\begin{equation}
G_a = -i\int dx^\perp dx^- (\partial_- \varphi)_i
(t_a)_{ij}\varphi_j \ .
\end{equation}
Note that the weak-isospin charges and the currents corresponding
to $G_a$  are not conserved if the zero mode $\omega_i$ is nonzero
since the cross terms in $\varphi,$ and $\omega$ are missing. Thus
$[H_{LF},G_a] \ne 0.$ Nevertheless, the charges annihilate the
vacuum: $G_a|0>_{LF}=0,$ since the dynamical fields $\varphi_i$
have no support on the LF vacuum, and all quanta have positive
$k^+.$ Thus the LF vacuum remains equal to the perturbative
vacuum; it is unaffected by the occurrence of spontaneous symmetry
breaking.

In effect one can interpret the $k^+=0$ zero mode field $\omega_i$
as an $x^-$-independent external field, analogous to an applied
constant electric or magnetic field in atomic physics.  In this
interpretation, the zero mode is  a remnant of a Higgs field which
persists from early cosmology; the LF vacuum however remains
unchanged and unbroken.

\section{Discretized Light-Front Quantization}

If one imposes periodic boundary conditions in $x^- = t + z/c$,
then the plus momenta become discrete: $k^+_i = {2\pi \over L}
n_i, P^+ = {2\pi\over L} K$, where $\sum_i n_i = K$
\cite{Maskawa:1975ky,Pauli:1985pv}.  For a given ``harmonic
resolution" $K$, there are only a finite number of ways positive
integers $n_i$ can sum to a positive integer $K$.  Thus at a given
$K$, the dimension of the resulting light-front Fock state
representation of the bound state is rendered finite without
violating Lorentz invariance.  The eigensolutions of a quantum
field theory, both the bound states and continuum solutions, can
then be found by numerically diagonalizing a frame-independent
light-front Hamiltonian $H_{LC}$ on a finite and discrete
momentum-space Fock basis.  Solving a quantum field theory at
fixed light-front time $\tau$ thus can be formulated as a
relativistic extension of Heisenberg's matrix mechanics.  The
continuum limit is reached for $K \to \infty.$ This formulation of
the non-perturbative light-front quantization problem is called
``discretized light-cone quantization" (DLCQ)~\cite{Pauli:1985pv}.
The method preserves the frame-independence of the Front form.

The DLCQ method has been used extensively for solving one-space
and one-time theories~\cite{Brodsky:1997de}, including
applications to supersymmetric quantum field
theories~\cite{Matsumura:1995kw} and specific tests of the
Maldacena conjecture~\cite{Hiller:2001mh}.  There has been
progress in systematically developing the computation and
renormalization methods needed to make DLCQ viable for QCD in
physical spacetime.  For example, John Hiller, Gary McCartor, and
I~\cite{Brodsky:2001ja,Brodsky:2001tp,Brodsky:2002tp} have shown
how DLCQ can be used to solve 3+1 theories despite the large
numbers of degrees of freedom needed to enumerate the Fock basis.
A key feature of our work is the introduction of Pauli Villars
fields to regulate the UV divergences and perform renormalization
while preserving the frame-independence of the theory.  A recent
application of DLCQ to a 3+1 quantum field theory with Yukawa
interactions is given in Ref.~\cite{Brodsky:2001ja}. One can also
define a truncated theory by eliminating the higher Fock states in
favor of an effective
potential~\cite{Pauli:2001vi,Pauli:2001np,Frederico:2002vs}.
Spontaneous symmetry breaking and other nonperturbative effects
associated with the instant-time vacuum are hidden in dynamical or
constrained zero modes on the light-front.  An introduction is
given in Refs.~\cite{McCartor:hj,Yamawaki:1998cy}. A review of
DLCQ and its applications  is given in Ref. \cite{Brodsky:1997de}.

The pion distribution amplitude has been computed  using a
combination of the discretized DLCQ method for the $x^-$ and $x^+$
light-front coordinates with a spatial lattice
\cite{Bardeen:1979xx,Dalley:2001gj,Dalley:1998bj,Burkardt:2001mf}
in the transverse directions.  A finite lattice spacing $a$ can be
used by choosing the parameters of the effective theory in a
region of renormalization group stability to respect the required
gauge, Poincar\'e, chiral, and continuum symmetries.

Dyson-Schwinger models \cite{Hecht:2000xa} of hadronic
Bethe-Salpeter wavefunctions can also be used to predict
light-front wavefunctions and hadron distribution amplitudes by
integrating over the relative $k^-$ momentum.

\section{A Light-Front Event Amplitude Generator}

The light-front formalism can be used as an ``event amplitude
generator" for high energy physics reactions where each particle's
final state is completely labelled in momentum, helicity, and
phase.  The application of the light-front time evolution operator
$P^-$ to an initial state systematically generates the tree and
virtual loop graphs of the $T$-matrix in light-front time-ordered
perturbation theory in light-cone gauge.  Given the interactions of the
light-front interaction Hamiltonian, any amplitude in QCD and the
electroweak theory can be computed.  For example, this method can  be used
to automatically compute the tree diagram hard amplitudes
$T_H$ needed to for computing hard scattering amplitudes such as the
deuteron form factor or $pp$ elastic scattering.

At higher orders, loop integrals only involve integrations
over the momenta of physical quanta and physical phase space $\prod
d^2k_{\perp i} d k^+_i$.  Renormalized amplitudes can be explicitly
constructed by subtracting from the divergent loops amplitudes with
nearly identical integrands corresponding to the contribution of the
relevant mass and coupling counter terms (the ``alternating denominator
method")~\cite{Brodsky:1973kb}.  The natural renormalization
scheme to use for defining the coupling in the event amplitude
generator is a physical effective charge such as the pinch
scheme~\cite{Cornwall:1989gv}.  The argument of the coupling is
then unambiguous~\cite{Brodsky:1994eh}.  The DLCQ boundary
conditions can be used to discretize the phase space and limit the
number of contributing intermediate states without violating
Lorentz invariance.  Since one avoids dimensional regularization
and nonphysical ghost degrees of freedom, this method of
generating events at the amplitude level could provide a simple
but powerful tool for simulating events both in QCD and the
Standard Model.  Alternatively, one can construct the $T-$matrix
for scattering in QCD using light-front quantization
and the event amplitude generator; one can then probe its spectrum by
finding zeros of the resolvant.

\section*{Acknowledgments}
Work supported by the Department of Energy under contract number
DE-AC03-76SF00515.  This talk is in large part based on
collaborations with the late Professor Prem Srivastava, the
outstanding theorist from the Universidade do Estado de Rio de
Janeiro and  a key innovator of the light-front formalism.  I
thank Professor Beatriz Gay Ducati and the other organizers of
this meeting for their outstanding hospitality in Brazil.

\end{document}